\documentclass[10pt,prl,aps,twocolumn,superscriptaddress,showpacs,floatfix]{revtex4}
\usepackage{graphicx}

\usepackage{amssymb}

\begin{document}

\title{Antiferromagnetic to valence-bond-soild transitions in two-dimensional SU(N) Heisenberg 
models with multi-spin interactions}

\author{Jie Lou}
\affiliation{Department of Physics, Boston University, 590 Commonwealth Avenue, Boston, Massachusetts 02215}

\author{Anders W. Sandvik}
\affiliation{Department of Physics, Boston University, 590 Commonwealth Avenue, Boston, Massachusetts 02215}

\author{Naoki Kawashima}
\affiliation{Institute for Solid State Physics, The University of Tokyo, Kashiwa, Chiba, Japan}

\begin{abstract}
We study two-dimensional Heisenberg antiferromagnets with additional multi-spin interactions which can drive the system 
into a valence-bond solid state. For standard SU$(2)$ spins, we consider both four- and six-spin interactions. We find continuous 
quantum phase transitions with the same critical exponents. Extending the symmetry to SU$(N)$, we also find continuous transitions 
for $N=3$ and $4$. In addition, we also study quantitatively the cross-over of the order-parameter symmetry from Z$_4$ deep inside 
the valence-bond-solid phase to U$(1)$ as the phase transition is approached.
\end{abstract}

\date{\today}


\maketitle
Two-dimensional quantum spin system with non-magnetic ground states have been at the forefront of condensed matter physics for more
than two decades \cite{anderson,read,senthil,sachdev}. Frustrated system have been investigated intensly \cite{diep}, but large-scale 
unbiased computational studies of their ground states are not possible, due to the ``sign problems'' hampering quantum Monte Carlo 
(QMC) methods \cite{henelius}. It was recently realized that one prominent class of non-magnetic states---valence-bond solids 
(VBSs)---can be accessed also without frustration, by adding certain multi-spin interactions to the standard $S=1/2$ Heisenberg 
antiferromagnet \cite{sandvik1}. These models enable detailed QMC studies of the antiferromagnetic (AF) to VBS quantum phase 
transition. It has been argued that this transition is associated with spinon deconfinement (hence the term  deconfined 
quantum criticality) and should, due to subtle quantum interference effects, be continuous \cite{senthil}. This scenario violates 
the ``Landau rule'', according to which a direct transition between states breaking unrelated symmetries should be generically 
first-order.

The theory of deconfined quantum criticality has generated a great deal of interest, as well as controversy
\cite{sandvik1,melko,jiang,weihong,kuklov1,nogueira1,motrunich,kuklov2,nogueira2}. Numerical studies of a Heisenberg hamiltonian with
4-spin interactions are generally in good agreement with the theory, showing a continuous transition with dynamic exponent $z=1$, 
large spin correlation exponent $\eta_s$, and an emergent U(1) symmetry \cite{sandvik1,melko,jiang}. Arguments for a first-order 
transition have also been put forward \cite{kuklov1,kuklov2}, based on numerical studies of lattice versions of the field-theory 
proposed \cite{senthil} to capture the AF--VBS transition. Other, similar studies reach different conclusions, however \cite{motrunich}. 
Further studies are thus called for.

\begin{figure}
\centerline{\includegraphics[width=4cm, clip]{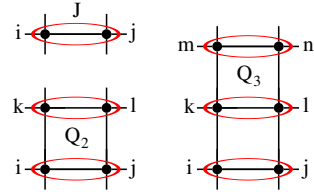}}
\vskip-2mm
\caption{(Color online) Interactions involving $p$ singlet projectors (illustrated by ovals enclosing two sites) on the square lattice. 
The two-spin ($p=1$) interaction $J$ is the Heisenberg exchange. Higher-order $Q_p$ terms with $p=2$ and $3$ are considered here. 
All translations and $90^\circ$ rotations of the site groupings shown here are included in the hamiltonian.}
\label{interactions}
\vskip-3mm
\end{figure}

In this Letter, we advance computational studies of the AF--VBS transition in two different ways. First, we consider the $S=1/2$
Heisenberg model including 4-spin and 6-spin interactions. The unperturbed Heisenberg model is defined by the hamiltonian
\begin{equation}
H_1 = J \sum_{\langle ij \rangle} {\bf S}_i \cdot {\bf S}_j = -J \sum_{\langle ij \rangle} C_{ij} + \frac{L^2J}{2},
\end{equation}
where $\langle ij \rangle$ denotes nearest neighbors on a periodic square lattice with $L^2$ sites and 
\begin{equation}
C_{ij}=\frac{1}{4}-{\bf S}_i \cdot {\bf S}_j
\label{cijdef}
\end{equation}
is the two-spin singlet projector. In the ``J-Q'' model introduced in 
\cite{sandvik1}, the following term is added to $H_1$;
\begin{equation}
H_2 = -Q_2 \sum_{\langle ijkl \rangle} \hskip-0.5mm C_{kl}C_{ij}.
\label{hamq2}
\end{equation}
The spin pairs $ij$ and $kl$ are located on adjacent corners of a 4-site plaquette, as illustrated in Fig.~\ref{interactions}. 
We denote the strength of the 4-spin term $Q_2$, with the subscript indicating two singlet projectors, and also consider a similar 
term with three stacked singlet projectors,
\begin{equation}
H_3 = -Q_3 \hskip-2mm \sum_{\langle ijklmn \rangle} \hskip-2.5mm C_{mn}C_{kl}C_{ij},
\label{hamq3}
\end{equation}
as also illustrated in Fig.~\ref{interactions}. Using an improved version \cite{sandvik2} of a ground-state QMC method 
operating in the valence-bond basis \cite{sandvik3}, we have studied the J-Q$_{\rm 2}$ and J-Q$_{\rm 3}$ models on lattices 
with $L$ up to $64$. We find critical AF--VBS points with the same set of exponents for both models, providing additional 
evidence of a universal deconfined critical point in this class of systems. 

In a second development, we have studied SU(N) symmetric versions of the J-Q$_{\rm 2}$ model, in the representation of the spin 
operators previously used in mean-field \cite{read} and QMC calculations \cite{kawashima} of the SU(N) Heisenberg model. We find 
continuous AF--VBS transitions also for $N=3$ and $4$ (whereas for $N>4$ the system is VBS ordered for all $Q_2>0$ \cite{kawashima,beach}). 

An open problem in previous studies of the J-Q$_{\rm 2}$ model was that the order parameter distribution inside the
VBS phase did not show the expected 4-fold symmetry. Instead, the distribution was always U(1) symmetric \cite{sandvik1,jiang}. An 
emergent U(1) symmetry close to criticality is indeed predicted by the field theory \cite{senthil} as a consequence of a
dangerously irrelevant operator, but deep inside the VBS phase the order parameter should exhibit Z$_{\rm 4}$  symmetry (which has 
been observed in other quantum models \cite{beach,lou1}). With the J-Q$_{\rm 3}$ model and the $N>2$ versions of the J-Q$_{\rm 2}$ 
model, we can now reach sufficiently deep inside the VBS phase to observe the expected U(1)--Z$_{\rm 4}$ cross-over. We present 
quantitative finite-size scaling results for the exponent governing the cross-over.

\begin{figure}
\centerline{\includegraphics[angle=0,width=6.75cm]{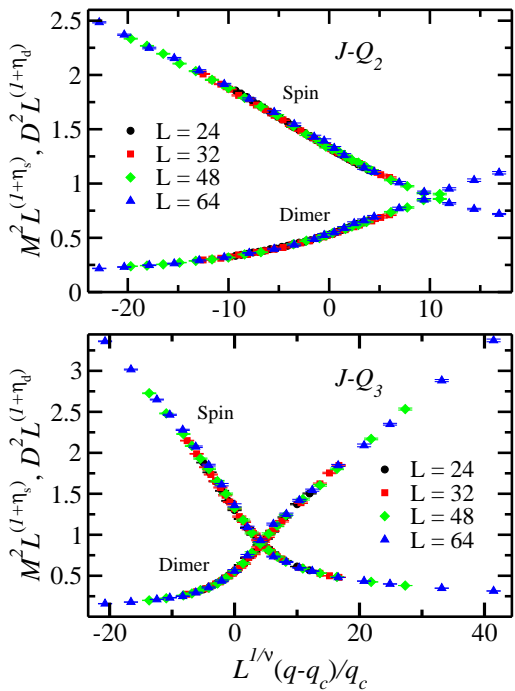}}
\vskip-2mm
\caption{(Color online) Finite-size scaling of the squared AF and VBS order parameters of the J-Q$_{\rm 2}$ and 
J-Q$_{\rm 3}$ models.} 
\label{fig:jq_dm}
\vskip-3mm
\end{figure}

For all the models, we compute the square of the staggered magnetization, $M^2=\langle {\bf M} \cdot {\bf M}\rangle$, where
\begin{equation}
{\bf M}=\frac{1}{L^2}\sum_{x,y}(-1)^{x+y} {\bf S}_{x,y}
\label{eqn:def_m}
\end{equation}
is the operator of the AF (spin) order parameter. We define the columnar VBS order parameter in terms of neartest-neighbor (dimer) 
correlators
\begin{eqnarray}
D_x=\frac{1}{L^2}\sum_{x,y}(-1)^{x} {\bf S}_{x,y} \cdot {\bf S}_{x+1,y},
\label{eqn:def_dx}
\end{eqnarray}
and $D_y$ defined analogously. We compute the square $D^2=\langle D_x^2+D_y^2\rangle$ and also study the probability distribution 
$P(D_x,D_y)$, with $D_x$ and $D_y$ evaluated in the configurations generated in the QMC sampling (as in \cite{sandvik1}).
To extract the critical points and exponents, we use standard finite-size scaling forms for the order parameters,
\begin{eqnarray}
M^2 &=& L^{1+\eta_s}F_s([q-q_c]L^{1/\nu}), \label{eqn:def_fs}  \\
D^2 &=& L^{1+\eta_d}F_d([q-q_c]L^{1/\nu}), \label{eqn:def_fd}
\end{eqnarray}
where $\eta_{s}$ and $\eta_{d}$ are the exponents governing the spin and dimer correlation functions, respectively, at criticality 
(the anomalous dimensions) and $1+\eta_{s,d}=2\beta_{s,d}/\nu$. Here we assume a dynamic exponent $z=1$, in accord with 
previous studies of the J-Q$_{\rm 2}$ model \cite{sandvik1,melko}, and use a single correlation length exponent $\nu$, as in the theory 
\cite{senthil}.

We first present results for the SU(2) models. Defining coupling ratios $q=Q_p/(J+Q_p)$, we find critical points $q_c=0.961(1)$ 
for $p=2$ and $q_c=0.600(5)$ for $p=3$. The former agrees with previous estimates \cite{sandvik1,melko,jiang}. Standard data 
collapse plots according to (\ref{eqn:def_fs})  and (\ref{eqn:def_fd}) are shown in Fig.~\ref{fig:jq_dm}. The critical exponents 
are listed on the first two lines of Table \ref{tab:sun}. Here it is very significant that all the exponents are the same for the two 
models. This supports the notion of a universal deconfined quantum critical point. Note that the order parameters decay as 
$L^{1+\eta_{s,d}}$ at the common critical point $q=q_c$. At a first-order transition, the order parameters should instead be size-independent 
at $q_c$, due to phase coexistence.

Comparing with previous results for the
J-Q$_{\rm 2}$ model, the results for smaller systems in \cite{sandvik1} were consistent with $\eta_s=\eta_d$ (with a value between those
found here), but the present results for larger systems clearly show that the spin and dimer exponents are different. The theory does not 
make any specific predictions for a relationship between $\eta_s$ and $\eta_d$, and they can be expected to be different. The exponents
$\eta_s$ and $\nu$ are in good agreement with values obtained using finite-temperature scaling in \cite{melko} (where $\eta_d$ was
not determined).

\begin{table}
\begin{center}
  \begin{tabular}{|cc|c| c |c |c|}
    \hline
    model, & symmetry &$\eta_s$  &$\eta_d$  &$\nu$ &$a_4$ \\ \hline
    J-Q$_{\rm 2}$, & $SU(2)$ &  0.35(2)  &  0.20(2)  &  0.67(1) & ---  \\ 
    J-Q$_{\rm 3}$, & $SU(2)$ &  0.33(2)  &  0.20(2)  &  0.69(2) &1.20(5)  \\ \hline
    J-Q$_{\rm 2}$, & $SU(3)$ &  0.38(3)  &  0.42(3)  &  0.65(3) &1.6(2) \\ \hline
    J-Q$_{\rm 2}$, & $SU(4)$ &  0.42(5)  &  0.64(5)  &  0.70(2) &1.5(2) \\ \hline
\end{tabular}
\end{center}
\vskip-3mm
\caption{Critical exponent for all the models studied. The cross-over exponent $a_4$ cannot be determined for the 
SU(2) J-Q$_{\rm 2}$ model, because no cross-over is observed for $L \le 64$.}      
\label{tab:sun} 
\vskip-3mm
\end{table}

\begin{figure}
\centerline{\includegraphics[angle=0,width=7cm]{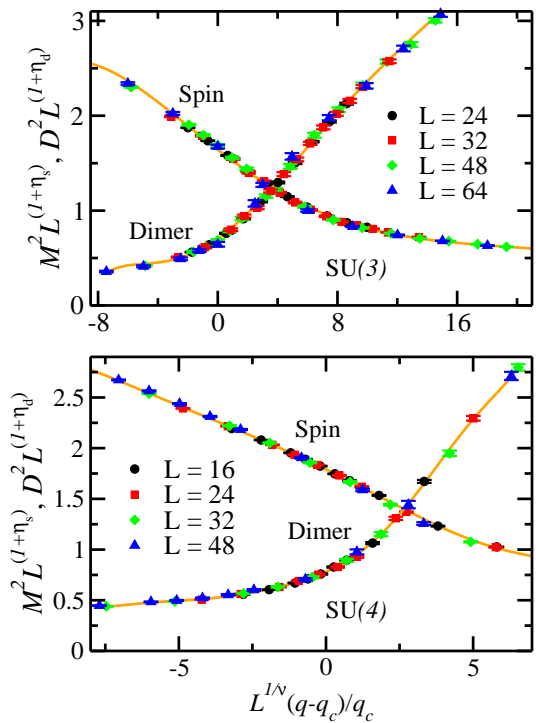}}
\vskip-2mm
\caption{(Color online) Scaling of the spin and dimer order parameters of the SU$(3)$ and
SU$(4)$ J-Q$_{\rm 2}$ models.}
\label{fig:su34_dmfs}
\vskip-3mm
\end{figure}

Next, we discuss the J-Q$_{\rm 2}$ model generalized to SU$(N)$ spins. Considering first the Heisenberg model, the hamiltonian 
can be written as
\begin{eqnarray}
H_{{\rm SU}(N)}=\frac{J}{N} \sum_{\langle ij\rangle}{{\bf S}_{i}^{\alpha\beta}{\bf S}_{j}^{\beta\alpha}} = - J\sum_{\langle ij\rangle}C_{ij}+\frac{2JL^2}{N^2},
\end{eqnarray}
where ${\bf S}_{i}^{\alpha\beta}$ is the generator of the SU$(N)$ algebra, with $\alpha,\beta=1,2,\cdots,N$ the different ``colors'',
and $C_{ij}$ is the generalization of Eq.~(\ref{cijdef}) to SU$(N)$. As in \cite{kawashima}, we focus on the simplest case, where the spins 
on sublattice A are expressed in the fundamental representation (i.e., with a single-box Young tableau). Spins on sublattice B are SU$(N)$ 
conjugates (dual representation) of those on A (a Young tableau with one column and $N-1$ rows). The states in this representation can be 
written in terms of permutations $P$ of the boxes, with 
\begin{equation}
|\overline{\alpha}\rangle_j\equiv \frac{1}{\sqrt{(N-1)!}}\sum_{P}(-1)^P|P(2)P(3)\cdots P(N)\rangle_j,
\end{equation}
with $\alpha=1,2,\cdots,N$ and $P(1)=\alpha$. An SU$(N)$ singlet of spins $i$ and $j$ on different sublattices is given by
\begin{eqnarray}
|{\rm singlet}\rangle_{ij}\equiv \frac{1}{\sqrt{N}}\sum_{\alpha=1}^{N}|\alpha\rangle_i \otimes |\overline{\alpha}\rangle_j.
\end{eqnarray}
QMC algorithms using these SU$(N)$ spins in the valence-bond basis are simple generalizations of the SU$(2)$ case 
\cite{sandvik2,sandvik3,kevin2}.
Instead of spins $\uparrow$ and $\downarrow$ for SU$(2)$, there are $N$ colors, and, thus, $N$ states of the 
space-time loops in the loop-algorithm \cite{sandvik2}. The off-diagonal matrix elements of the singlet projection operators are $1/N$ 
instead of $1/2$, and the overlap of two valence-bond states is generalized to $N^{n_\circ-L^2/2}$, where $n_\circ$ is the number of loops 
in the transposition graph. The 4- and 6-spin terms (\ref{hamq2}) and (\ref{hamq3}) are written explicitly using products of singlet 
projectors and have obvious generalizations to SU$(N)$.

Our results for the SU$(3)$ and SU$(4)$ versions of the J-Q$_{\rm 2}$ model are consistent with 
continuous AF--VBS critical points, with no signs of first-order behavior. The critical points are $q_c=0.335(2)$ and 
$q_c=0.082(2)$ for $N=3$ and $4$, respectively. Scaling plots giving the critical exponents are shown in Fig.~\ref{fig:su34_dmfs} and  
numerical values are listed in Table \ref{tab:sun}. It can be noted that, as a function of $N$,  $\nu$ does not change appreciably, 
$\eta_s$ increase slowly, and $\eta_d$ increases significantly. An increasing spin exponent is consistent with $\eta_s=1$ in the 
$N \to \infty$ theory \cite{senthil}. As already noted, there are no predictions for $\eta_d$. We could, in principle, consider still 
higher $N$, but with $J>0$ the system is always in the VBS state for $N=5$ and higher \cite{kawashima,beach}. A transition could 
presumably be reached for $J<0$, but this causes QMC sign problems.

\begin{figure}
\centerline{\includegraphics[angle=0,width=5.75cm]{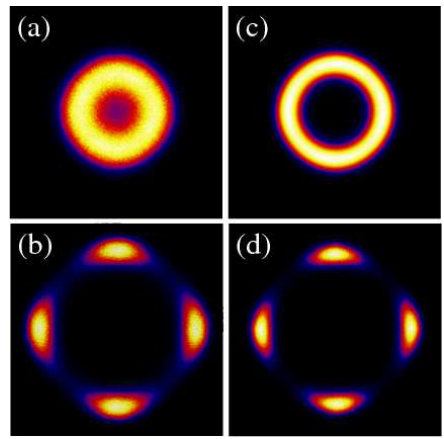}}
\vskip-2mm
\caption{(Color online) Dimer order distribution $P(D_x,D_y)$ for $L=32$ systems. The left panels are for the J-Q$_3$ model at 
$q=0.635$ (a) and $q=0.85$ (b), and the right panels are for the SU$(3)$ J-Q$_2$ model at $q=0.45$ (c) and  $q=0.65$ (d).}
\label{fig:hist}
\vskip-3mm
\end{figure}

The dimer order distribution $P(D_x,D_y)$ can be used to investigate the VBS order parameter symmetry \cite{sandvik1,kawashima}. 
As shown in Fig.~\ref{fig:hist}, for large $q$ the robust VBSs in the SU$(2)$ J-Q$_3$ model and the SU$(3)$ and SU$(4)$ versions of 
the J-Q$_2$ model result in histograms with clearly visible columnar Z$_4$ features (i.e., peaks on the $D_x$ and $D_y$ axis, as opposed 
to $45^\circ$ degree rotated histograms expected for a plaquette state). However, in the SU$(2)$ J-Q$_2$ model the histograms are
ring-shaped for all system sizes currently accessible, even in the extreme case of $q=1$ ($J=0$). In all cases, we see U$(1)$ symmetric 
histograms as the critical point is approached, in agreement with one of the salient features of deconfined quantum criticality 
\cite{senthil}. 

\begin{figure}[t]
\centerline{\includegraphics[width=6.75cm,clip]{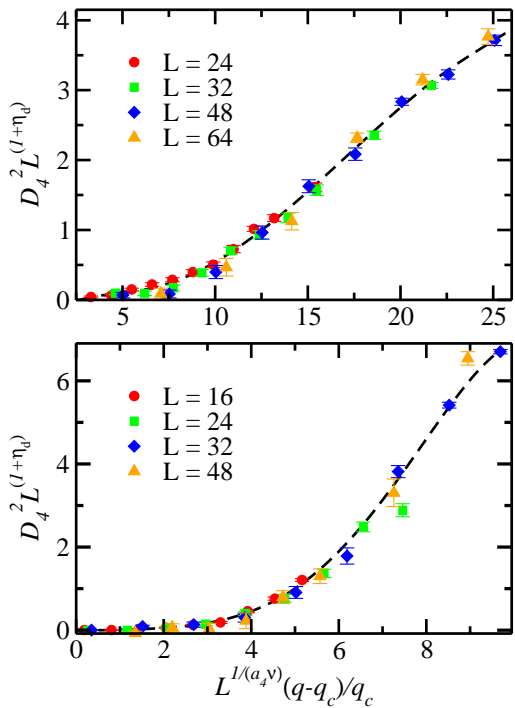}}
\vskip-3mm
\caption{(Color online) Finite-size scaling of the square of the anisotropic order parameter $D_4$ in the J-Q$_3$ model (upper panel) 
and SU$(3)$ J-Q$_2$ model (lower panel).}
\label{z4}
\vskip-3mm
\end{figure}

Defining an order parameter sensitive to the symmetry,
\begin{eqnarray}
D^2_4&=&\int dD_xdD_yP(D_x,D_y)(D_x^2+D_y^2)\cos(4\theta) \nonumber \\
   &=&\int dr\int_0^{2\pi}d\theta P(r,\theta)r^3\cos(4\theta),
\label{eqn:def_d4}
\end{eqnarray}
where $\theta$ is the angle corresponding to a point $(D_x,D_y)$, we proceed as in \cite{lou2} (which deals with a classical system with a 
dangerously irrelevant perturbation) to extract the exponent governing the length-scale $\Lambda$ of the Z$_4$--U$(1)$ cross-over (and the spinon 
confinement). Z$_4$ features should appear for $L > \Lambda$, which is predicated \cite{senthil} to scale as $\Lambda \sim \xi^{a_4}$ where $\xi$ 
is the correlation length and $a_4 > 1$. We analyze $D_4$ assuming the scaling form \cite{lou2};
\begin{equation}
D^2_4 = L^{1+\eta_d}F_4(qL^{1/a_4\nu}).
\label{eqn:d4fs}
\end{equation}
This form describes the cross-over, as shown in Fig.~\ref{z4} in two cases. The values of $a_4$ are listed in Table \ref{tab:sun}. 
The large error bars reflect slow evolution of the VBS angle in the QMC simulations. It is nevertheless clear that $a_4>1$ (and
increasing with $N$), reflecting emergent U$(1)$ symmetry due to a dangerously irrelevant perturbation.

The results presented here support deconfined quantum criticality. Although one can still, 
in principle, not completely rule out very weakly first-order transitions based on these calculations, the universal behavior for the
two SU$(2)$ models makes this less likely. The common exponents for the J-Q$_{\rm 2}$ and J-Q$_{\rm 3}$ models at the very least suggest 
close proximity to a universal critical point. The detailed information now available from QMC simulations should be useful to further 
advance the theory.

In a very interesting experimental development, Itou {\it et al.} have recently measured the spin-lattice relaxation rate $1/T_1$ in a 
layered organic compound which seems to be near-critical \cite{itou}. It has been argued that, in spite of the triangular lattice, the AF--VBS 
transition in this kind of system should be in the same class of deconfined quantum critical points discussed here \cite{xu}. The exponent 
$\eta_s$ governs the temperature scaling of $1/T_1$, and the value $\eta_s\approx 0.35$ is in excellent agreement with the experiment over a 
wide range of temperatures. Further experiments should elucidate the nature of the ground state and whether it is indeed close to a deconfined 
quantum-critical point.

{\it Acknowledgements}---We would like to thank R.~Kaul and A. Vishwanath for useful discussions. This work was supported by NSF grant No.~DMR-0803510 
(J.~L. and A.~W.~S) and MEXT Grant-in-Aid for Scientific Research [(B)(19340109) and Priority Area ``Novel States of Matter Induced by 
Frustration'' (19052004)] (N.~K.).

\null


\begin{thebibliography}{00}
\bibitem{kevin2}
An equivalent way to generalize to $N>2$ is to consider spin $S=(N-1)/2$ models including higher-order exchange 
operators rendering the hamiltonian SU$(N)$ symmetric. This approach was developed for the Heisenberg model in \cite{beach}.
\end{thebibliography}
\end{document}